\documentclass{article}
\usepackage{graphicx} 

\title{Relational Observables, Quiddities, and Structural Realism} 
\date{\today}

\author{Emily Adlam  \thanks{Philosophy Department and Institute for Quantum Studies, Chapman University, Orange, CA92866, USA \texttt{eadlam90@gmail.com} }}

   \usepackage[top=2.5cm, bottom=2.5cm, left=2.5cm, right=2.5cm]{geometry}
 \usepackage{setspace}
\usepackage{url}
 \usepackage{natbib}
\date{\today} 
\setcitestyle{round}

\begin{document}

\maketitle

\begin{abstract}

  In this article, I argue that modern spacetime physics causes problems for a number of traditional accounts of modality, but also offers important new ideas about the connection between modal and non-modal features of reality. I suggest that  recent work on relational observables in general relativity and quantum gravity can help us understand how non-modal features of reality could arise from modal features of reality within some form of modal ontic structural realism. In particular, I argue that the notion of an `internal view,' as  employed in the partial/complete observables formalism and the quantum reference frame formalism, is an important conceptual insight which can help address outstanding problems in the philosophy of lawhood and modality.

 \end{abstract}

\section{Introduction}

Spacetime plays an important role in many philosophical accounts of laws and modality. But our understanding of the nature of spacetime has undergone significant revisions in the context of modern physics. This leads to difficulties for a number of traditional accounts of modality, but on the other hand, conceptual innovations in spacetime physics may offer valuable new ideas on this topic.  In this article, I will argue the the partial/complete observables formalism as applied in general relativity and quantum gravity suggests an interesting novel way of thinking about the connection between modal and non-modal features of the world.

I begin in section \ref{keepdistinct} by  distinguishing between `contingentist' and `structuralist' conceptions of  the connection between modal and non-modal features of reality; I argue that   modern spacetime physics causes significant problems for contingentist approaches, particularly if we want to avoid positing quiddities. Thus I suggest that  these considerations motivate the adoption of a structuralist approach, such as some form of modal structural realism in which non-modal features arise from qualitative modal properties rather than vice versa.

In section \ref{partial} I will describe some recent developments in our understanding of diffeomorphism-invariant spacetime theories and quantum gravity, including the partial/observables approach, and I will argue that these formalisms can be understood as instantiating some form of modal ontic structural realism. Finally in section \ref{modal} I will explain how these formalisms make use of the notion of an `internal view,' and I will argue that this idea offers a useful insight into the emergence of non-modal features of reality from modal features. Thus this idea may be used to  formulate a viable form of modal structural realism which can account for the relations between modal and non-modal features of reality in the context of modern spacetime physics.

\subsection{Methodological Note \label{methodology}}

In this article I will largely  adopt a metaphysically naturalistic methodology, i.e. I will assume that our metaphysics should be informed by our best scientific theories. Thus, for example, I will argue that the conceptual innovations inherent in general relativity and quantum gravity are a sufficient motivation for investigating whether the metaphysics of modality can be formulated in a way which does not rely on giving a privileged role to spatiotemporal features of reality.

Now, it should be emphasized that several of the formalisms I am discussing here, such as the partial/complete observables program and the Page-Wootters approach,  are still active research programs; the science here is not fully settled. So  although I will make use of these formalisms here, it is not my intention  to argue that they are conclusive evidence in favor of any particular philosophical view, and my approach does not rely on the assumption that these formalisms are completely correct. Rather, the point I want to make is that regardless of whether or not these formalisms \emph{are} in fact correct, they represent interesting conceptual innovations which may give us useful new ideas about how to address problems in metaphysics in ways which reflect recent developments in physics. In particular, I will argue  that the notion of an `internal view' as it is used in these  formalisms offers a helpful way of conceptualizing the emergence of non-modal features from modal features, and this is a valuable insight regardless of the ultimate fate of the specific formalisms in which it appears.

 Now of course when pursuing a metaphysically naturalistic approach it comes with the territory that the resulting metaphysics will be limited in scope, so the approach suggested here will be suitable for worlds with modal structure that shares some relevant features with the partial/complete observables formalism or the QRF formalism, but not  necessarily for all possible worlds. Therefore this approach is incompatible with some very general  ways of thinking about the relations between modal and non-modal features of reality. For example, \cite{Schaffer2005-SCHQK-2} suggests a  characterization of this issue based on the idea that `the question of the relation between properties and their powers is transformed into the question: which possible worlds should one countenance?' But I will not be able to understand the question in that way here, because I am suggesting an account of this relation which works only for certain  kinds of modal structures, and therefore I am committed to the claim that the nature of the relation between modal and non-modal structure may itself be different in different possible worlds with different laws, so any identification of the possible worlds to be countenanced would itself have to be indexed to a world with a particular set of laws. 
 
 I find this limitation reasonable, because I do not think it makes sense to expect that the relationship between modal and non-modal features of reality can be fully characterized in a way that is completely independent of the \emph{nature} of the modal structure concerned.  In particular, we will see throughout this article that spacetime plays an important role in various stories one might tell about the connection between modal and non-modal features of reality, and thus our account of these connections should  be at least somewhat responsive to what physics has to tell us about the nature of spacetime.  That is, the relationship between modal and non-modal features of reality is not a purely metaphysical matter; for at least some kinds of modal features it is in part a scientific question, just as the nature of spacetime is at least in part a scientific question.  This suggests that for some kinds of modal features, such as laws and causal profiles, the connection between these features and the non-modal features of reality may not hold as a matter of metaphysical necessity - it may look different in worlds with very different kinds of laws, so we should be willing to diverge from the possible worlds formalism in some cases.

\section{Connections Between Modal and Non-Modal Features \label{keepdistinct}}

In this article I will use the term `modal features of reality' to refer generally to laws and dynamics, as well as  causal profiles, modal structures, dispositions, symmetries, and other such features which are typically understood as falling outside of the realm of `actualities.'  By contrast, I will understand the `non-modal features of reality' to include  entities, categorical or occurrent properties, and non-modal relations such as relative distance. Throughout this paper I will focus exclusively on qualitative physical properties and relations, in both the modal and non-modal cases; the question of how these things are related to non-qualitative modal properties is interesting, but not something I will address here. Note also that one might reasonably worry that the distinction between modal and non-modal features is not entirely well-defined: I think this is right, and indeed I will ultimately argue for a position in which non-modal properties are simply views on modal structure, but  the distinction is at least well enough understood to make use of for now. 

Here I am interested in understanding how modal features of reality are related to non-modal features of reality. I will divide the set of possible approaches to this question into two classes:

\begin{itemize} 

\item \textbf{Contingentist:} there exist some non-modal properties or relations which are defined independently from their modal or nomic role. 

\item \textbf{Structuralist:} all properties or relations are defined at least in part by their modal or nomic role. 

\end{itemize}

In this section, I will argue that modern spacetime theories pose a special problem for contingentism, because contingentists who want to avoid quiddities are naturally led to a position which gives spacetime a privileged role. I will then argue that structuralist approaches seem more promising in this regard, but face a number of important open questions. 

\subsection{Contingentist Approaches}

  Any way of thinking about the connection between modal and non-modal features of reality which seeks to derive the former from the latter is necessarily contingentist, since in such a picture the non-modal properties must be defined in advance of the modal features. For example, one common contingentist approach is the Humean picture, which suggests that laws (and other modal/nomic features) are to be identified as systematizations of patterns in a Humean mosaic, which consists of a distribution of categorical properties over spacetime \citep{doi:https://doi.org/10.1002/9781118398593.ch17}. Evidently the properties making up the mosaic must be  definable in advance of any modal features, so the  identities of these properties  cannot depend on their modal or nomic features.  

There are also a number of contingentist approaches which treat modal and non-modal features as separate classes such that neither can be derived from the other. For example, in the universals picture, laws of nature   are defined as contingent relations between universals \citep{Armstrong1983-ARMWIA,10.2307/187350}, which means that the set of universals must be defined separately from the laws, since  a relation between a pair of universals cannot be contingent if the definition of the universals depends on that relation. Similarly,  in the recent literature on constraint-based views  of lawhood \citep{adlam2021laws,chen2021governing, Meacham2023-CHRTNL},  laws are  seen as governing in virtue of narrowing down the space of possible courses of history, and thus  the properties   featuring in the space of possible courses of history must be defined separately from the laws. Thus if we assume  that in these models the laws exhaust the modal features of reality, it is clear that such models are committed to the existence of at least some properties which can be defined independently from their modal or nomic features.

Many existing contingentist approaches give a special role to spatiotemporal properties and relations. We can understand the reasons for this by noting that contingentist approaches face the following difficulty. If there exists more than one type of fundamental non-modal property or relation, we will need some way of individuating these  properties or relations. And in the contingentist picture, we cannot individuate them by appeal to their modal or nomic features. Thus it seems likely that we will be forced to individuate them by appeal to \emph{quiddities}: intrinsic essences which are independent from any facts about their function or behaviour. 

However, there are a number of serious philosophical objections to quiddities \citep{Bird2007-BIRNML-2, Lierse1996-LIETJH,Black2000-BLAAQ, Shoemaker1998-SHOCAM}. For example, they lead to variety of  epistemic problems:  we cannot access quiddities directly and thus if two properties with distinct quiddities were to share a causal profile, we would never be able to distinguish between them \citep{Berenstain2016-BERWAS-2}. Quiddities also do not seem to be a very good fit with the way in which properties are actually defined in science \citep{Blackburn1990-BLAFIS,Ellis1994-ELLDE}. Thus there are clear motivations to try to formulate an account of the relation between modal and non-modal features of reality without appeal to quiddities. 

While it is in principle possible that there exists some third way of individuating properties which does not involve appeal to either quiddities or modal features, certainly these are the most commonly discussed approaches. Thus unless some third way can be found, it follows that if we want to adopt a contingentist approach without invoking quiddities, we will need to insist that there exists only \emph{one} type of fundamental non-modal property or relation, so there is no need for individuation.  What kind of property or relation could play this role? 

Most extant views of this kind agree that it should be spatiotemporal in nature. For example, Super-Humeanism is a contingentist view which postulates  a minimal ontology consisting  only of massless matter points and relative distance relations between them \citep{pittphilsci14095}, so  relative distance is the only fundamental non-modal relation. Similarly, position-first Humeanism regards position as the only fundamental property \citep{doi:https://doi.org/10.1002/9781118398593.ch17}. On the non-Humean side, dispositional essentialism \citep{Bird2005-BIRLAE} can be regarded as a contingentist view if we posit a basic ontology consisting of objects possessed of modal properties, where the assignation of properties to objects is a contingent matter. But before we can assign modal properties to these objects we will have to individuate them, and if we are not using quiddities then we cannot individuate them by appeal to non-modal properties, so they are commonly individuated by appeal to their spacetime positions and/or the spatiotemporal relations between them. Moreover, time in particular plays a special role in the dispositional picture, since it is treated as a `fundamental dimension ... in which the disposition or powers manifest themselves, producing their effects' \citep{Lam_2023}.  So again in this picture   spatiotemporal properties or relations play the role of the specially privileged fundamental non-modal property or relation.

These approaches are  predicated on the idea that there is something special about spatiotemporal properties which justifies giving them a sui generis status: as  \cite{Bhogal2021-BHOHNE} put it, `It’s intuitive to claim that nothing can be positively charged if it doesn’t repel other things that are positively charged. But things look different with position.' This is a common intuition, and thus these quiddity-free contingentist approaches follow in a long tradition of privileging spacetime in our attempts to describe fundamental reality. Given the argument above, we can see that this shared feature is not incidental: there are good reasons to think that any contingentist view without quiddities will have to pick some kind of privileged property or relation, and spatiotemporal properties are the obvious choice. 

However, there are some reasons to be cautious about this. First of all,  many of the arguments for structuralism about properties and relations appear to carry over to spacetime, as emphasized by  spacetime functionalists: `one cannot simply posit that a certain structure, or type of structure, is spatiotemporal. Rather, a structure may only count as spatiotemporal in virtue of playing the right sort of role in the laws of a theory' \citep{pittphilsci15860}. Whether or not one fully endorses the spacetime functionalist programme, the work done by spacetime functionalists serves to highlight the deep connections between spatiotemporal notions and dynamical features of our theories, thus raising questions about whether it is reasonable to treat spatiotemporal properties or relations as sui generis in the way the contingentist strategy demands.  In particular, if the mandate to get rid of quiddities is partly motivated by the view that we should be guided by scientific practice and its functionalist approach to the definition of properties, then one might think that we should also be guided by the implicit functionalism about \emph{spacetime} that is evident in some aspects of scientific practice, and hence we should  allow that spatiotemporal properties and relations may also be defined by their dynamical role.

Moreover, this line of argument becomes significantly more compelling when we move to theories with a dynamical spacetime. For example, in General Relativity  (GR) relative distances are determined dynamically by the distribution of matter in spacetime, since distance is a metrical quantity and the metric is dynamical in GR. This is true both for relative spatial distance in a particular choice of reference frame, and for relative spatiotemporal distance. So in such a theory, it looks much less reasonable to define `relative distances' prior to saying anything about the connection of relative distance to the dynamics: since spacetime is dynamical, then if other properties and relations are to be defined in terms of their dynamical role, it seems natural to think that should be the case for the spatiotemporal properties and relations as well. 

To be clear, the point here is not that GR inevitably forces us to define understand distances in terms of their dynamical role: in principle it is still possible to treat the relative distances as something like a fundamental primitive, and then make sense of  GR by simply describing how these distances are affected by distributions of matter. For example, this is the approach taken by  Super-Humeanism: `Humeanism can accommodate a dynamical space-time in the following manner: considering an initial configuration consisting in metrical relations between space-time points and some of these points being occupied by primitive stuff, the further development of the metrical relations depends on how the stuff is distributed in that initial configuration' \citep{7c3bdfa2-a379-30cd-8b44-fc7327bf4c2b}.  However, note that in this approach, the dynamical nature of spacetime is just a contingent fact about the structure of the best systematization of  the relative distance relations of the fundamental particles. Had the development of the relative distance relations over time been different, we would have arrived at a different best system, and the equations of this best system might not have resulted in a dynamical spacetime. So although this approach may work at a technical level, one may still worry that it does not do justice to the insight that General Relativity gives us into the nature of spacetime: if we wish to take seriously the conceptual revolutions suggested by GR, then we might think that the dynamical nature of spacetime should be \emph{essential} to it, not merely a contingent feature of the actual configurations of relative distance relations. 

And indeed, one important reason to take these conceptual insights seriously is that they are likely to  become increasingly important as we advance beyond General Relativity. For example,  a number of approaches to quantum gravity suggest  that there is no spacetime at a fundamental level at all \citep{Lam_2023,cc,Huggett_2013}; and if spacetime is not fundamental, it probably does not make sense to posit that there exists a specially privileged fundamental property or relation which is  spatiotemporal in nature. So although it may be possible at a technical level to accommodate dynamical spacetimes within a contingentist view like Super-Humeanism, there is a definite possibility that the underlying conceptual tension will evolve into an outright incompatibility as we move towards quantum gravity. 
 
Of course, contingentists might hold out hope that in the context of a theory of quantum gravity, the specially privileged spatiotemporal property or relation can simply be replaced with some kind of \emph{non}-spatiotemporal property or relation instantiated in the relevant theory  - for example, entanglement \citep{Jaksland2020-JAKEAT-2}, or the adjacency relations appearing in  Loop Quantum Gravity \citep{cc}. But the difficulty is that most of the other possible backgrounds we might appeal to seem to have some modal content already built in: as Lam and W\"{u}trich put it, `In the two examples above, already the kinematical structures—the causal sets or the spin networks—are subject to axiomatic demands by the theory and so are steeped in laws' \citep{Lam_2023}. That is, any non-spatiotemporal property or relation we might appeal to will likely already be somewhat modal in nature, so it will not be suitable to play the role of the specially privileged \emph{non-modal} property or relation which the contingentist appears to require.\footnote{Of course, in a sense this is true of spacetime as well, but perhaps due to our familiarity with spatiotemporal notions it seems easier to think of spacetime as something like a container rather than a set of modal constraints on the possibilities for recombination. For a different kind of space it seems much harder to see the space as non-modal, and  therefore the possibility of defining patterns in some kind of non-spatiotemporal arena does not avoid the problem posed by dynamical spacetimes.}

\subsection{Structuralist Approaches \label{structuralist}} 

It seems difficult to make a contingentist approach work without either using quiddities or giving spacetime a specially privileged role. This suggests that in the context of modern physics, we may be better served by the adoption of some kind of structuralist approach, in which \emph{all} properties and relations are ultimately to be defined in structuralist terms. 

If we adopt this approach, we are probably going to be pushed towards something in the vicinity of  modal ontic structural realism, where we take it that only modal structure is fundamental, and non-modal features of reality are defined by this fundamental modal structure. This might involve some kind of eliminativism about non-modal features, or an emergentist view in which non-modal features emerge from the fundamental modal features in certain regimes, or  a modal version of `Moderate Structural Realism' as formulated by \cite{Esfeld2011-ESFOSR}, in which  properties and relations are `on the same ontological footing, being given ``at once” in the sense that they are mutually ontologically dependent on each other.'  Since these approaches allow that non-modal properties and relations can be individuated by their modal features, they can  avoid quiddities without giving spatiotemporal properties or relations a privileged role. 

However, views of this kind face a number of well-known challenges. First of all, there is the problem of accounting for the manifest image: our immediate experience of the world seems to reveal that it is filled with entities in possession of categorical properties\footnote{As pointed out by an anonymous reviewer, some dispositionalists might dispute this claim on the grounds that it is at least conceptually possible that all properties are really dispositional in nature. And in fact that is exactly the kind of view I will also advocate in this paper; so my point here is not to say that it is impossible to understand our experience in other ways, but simply to note that for many people it seems highly intuitive to say that   entities  possess intrinsic properties of their own, so there is an onus on any view which denies the reality or fundamentality of categorical properties to offer some explanation for this feature of our experience.}, and how can we make sense of that if the basic ontology is entirely modal?  Regardless of whether the non-modal features emerge, are eliminated, or are understood as aspects of the modal structure, something is going to have to be said about how we come to have experiences \emph{as of} non-modal features, and this seems challenging. 
 
Such worries are connected to a common objection against structuralism, based on the idea that it is not coherent to posit `relations without relata.' This is a particularly pressing concern for an eliminativist approach, but even in the emergentist approach one might worry that it doesn't make sense to posit  modal structures \emph{prior} to the objects that are supposed to instantiate them.  As \cite{McKenzie2017-MCKOSR} puts it, `if relations  are  – at least in the entanglement and  spatiotemporal cases – relations between objects, it seems we cannot	 say	 that there exist the former and not the latter  without  resorting to  a	 perilous  Platonism.' A popular response here is that we should think of objects as merely `points of intersection' between relations \citep{Cassirer1956-CASDAI-5,Eddington1928-EDDTNO-3}, but this seems more challenging in the case where the structures are modal. In particular, if one thinks of modal structure as characterizing possibilities or as in some sense constructed from possibilities, then  it is perhaps not obvious how the intersection of two relations constructed from or pertaining to possibilities could bring into being an \emph{actual} object, since one might naturally think that non-actual things cannot compose or subvene actual things. 
 
Similar worries arise in the context of  Moderate Structural Realism when we try to apply it specifically to modal structures. Perhaps we can make sense of the idea that objects and non-modal relations are different aspects of one another, but it seems harder to conceptualize the idea that modal and non-modal features of reality could be just different aspects of one another.   \cite{Esfeld2011-ESFOSR} suggest an approach inspired by Spinoza's view of properties as modes of existence \citep{Spinoza1677-SPIE}, where we say that `a structure consists in objects whose ways of existence are certain relations
that they bear to each other,' and  with respect to modal relations they suggest that `there are concrete physical relations as the ways in which the fundamental physical objects exist,' and that  these relations are `powers or dispositions to bring about certain physical effects.' But even if we agree that having the power to bring about certain physical effects is a mode of existence, one might think that in order to define the relevant physical effects we will first need to define some non-modal properties.  For example, \cite{Esfeld2011-ESFOSR} suggest that in the GRW interpretation of quantum mechanics, an entangled state is `the power to bring about classical physical properties that have a rather precise value through spontaneous localizations.' Yet surely in order to make sense of this we will first need to define the class of classical physical properties which can be brought about by this power, and thus we will need to define some set of  non-modal properties before we start defining the structure, and that seems to lead us back down the road towards requiring quiddities for these properties. 

So although some kind of modal ontic structural realism seems like a promising way of understanding the relation between modal and non-modal features of reality in the context of modern spacetime physics, there are still open questions about the nature of the connection between modal and non-modal features in this kind of picture, particularly with regard to the plausibility of actual properties and relations emerging from a \emph{purely} modal fundamental reality. I will now argue that recent innovations in the study of relational observables in GR and quantum gravity may offer some useful insights about how this kind of structuralist view can be made to work.

\section{Diffeomorphism-Invariant Theories\label{partial}}

Consider a theory $T$ which has diffeomorphisms as a local symmetry, such as GR. That is, if we take any history which is dynamically possible according to $T$, and apply a local diffeomorphism to  some spacetime region, we will end up with another history  which has identical  initial and final conditions and which is also dynamically possible according  $T$. This  seems to make the theory $T$ indeterministic in a particularly radical and troubling way, since the initial and final conditions together with the equations of $T$ will fix almost nothing about the intervening history. To avoid this conclusion, the standard move is to stipulate that histories related by a diffeomorphism transformation are  physically identical, so determinism is restored \citep{Wallacenew,Earman2002-EARTMM}.

But imposing diffeomorphism-invariance in this way has an interesting consequence:  it entails that variables which  `can be expressed as local functions of the coordinates'  \citep{rovelli2022philosophical} are not `observables,' which in this context is understood to mean that they are not physically real at all. For such variables are not invariant under diffeomorphisms, and if we wish to maintain that histories related by diffeomorphisms are physically identical,  we must insist that variables which can be changed by a diffeomorphism are not  physically real in and of themselves. \cite{Einsteindialog} thus writes that in GR, `the gravitational field at a certain place does not correspond to something `physically real,' but in connection with other data it does.'  
 
However, in our ordinary experience of the world we encounter many quantities that at least na\"{i}vely look like local functions of the coordinates - for example, when I observe the momentum of a baseball, one might naturally think that I am making an observation about the goings-on in some particular spacetime region. And indeed, much of classical and quantum physics is standardly expressed in terms of local functions of the coordinates. So there is a prima facie problem about how physically meaningful observables and regimes suited to classical and quantum physics could ever appear in a diffeomorphism-invariant theory. 

A variety of approaches exist, often based on the idea that the observables in a diffeorphism-invariant theory must ultimately be relational in character. Our ordinary experience of objects and categorical properties thus emerges as a consequence of the fact that we are part of the universe and so we are looking at these relational observables from the inside, so to speak; we have an \emph{internal view}. For example, the partial/complete observables formalism, originally proposed by  \cite{Rovelli_2002}, distinguishes between partial observables, which are physical quantities `to which we can associate a measuring procedure leading to a number'  \citep{Rovelli_2002}, and  complete observables, which are diffeomorphism-invariant and hence physically real. Rovelli argues that partial observables are not diffeomorphism-invariant and hence they are not physically real,  but the \emph{relations}  between partial observables are complete observables, and therefore such relations are individually real. So, the thought goes, although partial observables are not physically real, they can nonetheless be observed by a detector or an observer which is physically embodied, because the detector or observer is thereby observing a \emph{complete} relational  variable pertaining to the relation between the detector or observer itself and the entity it is measuring. As \cite{rovelli2022philosophical} put it: `The quantity measured by the detector is a local function of the metric, in the location determined by the detector. The full diffeomorphism invariance of the pure gravity dynamics, in other words, is physically broken by the detector itself being located somewhere.'  Thus as  \cite{pittphilsci4223} explains, `the dynamics is then spelt out in terms of relations between partial observables. Hence, the theory formulated in this way describes relative evolution of (gauge variant) variables as functions of each other.'

Now, it may seem counterintuitive to say that there can be a relation between two things which are not real, or that something which can be measured is not real. However, worries of this kind can perhaps be ameliorated by noting  that when a physicist says that some variable is not `physically real' (in Einstein's terminology) what is really meant by this is that it is not \emph{individually} real. That is, we cannot think of partial observables as existing on their own as individuals in the usual categorical sense, but we can perhaps still think of them as having some derivative form of reality in virtue of forming parts of the categorical observables, thus explaining how they can still stand in relations and be measured. 

Of course, it may still sound strange to suggest that the relata of a relation  exist only in virtue of the relation, but this is because the partial/complete observables formalism is specifically asking us to accept a reconceptualization of the nature of relations and relata: as  \cite{pittphilsci4223} puts it, `we talk about correlations in terms of quantities that are correlated,' so it seems natural to think that the correlations ontologically depend on the quantities, but that is simply not the case for complete observables: in this scenario `there are no independent subjects that are the ‘bearers’ of properties and the ‘enterers’ of relations.'  While this is certainly a revisionary approach to relations and relata, there are good reasons to take it seriously - after all, it is quite well-supported by our most advanced physical theories, and at least mathematically speaking we know how to formulate such a picture, since that is exactly what the standard formalism of GR gives us. The existence of  a  well-defined mathematical formulation of this way of thinking about relations provides at least some reason to belive that it is coherent, though certainly more  philosophical work is needed to clarify its metaphysics. 
  
In this connection, \cite{pittphilsci4223} has argued that the partial/complete observables framework is a good fit with something like ontic structural realism. In particular, he suggests that this formalism  shows us how to address the `no relations without relata' objection. For in the partial/complete observables framework, `the relations are the correlations ... (the gauge invariant, complete observables), and the ‘relata’ would be the non-gauge invariant, partial observables. But the partial observables being non-gauge invariant do not correspond to physical reality: only the complete observables do. Partial observables correspond to an arbitrary choice of gauge that can be transformed away.' That is, Rovelli's framework offers a concrete example of a model in which relations are fundamental and relata emerge from them: complete observables are relations, and what they relate are partial observables, but the  partial observables are not individually real and  thus the relata here are ontologically dependent on  the relations. Given that the partial/complete observables picture is mathematically coherent and supported by contemporary physics, this should lessen our worries about the coherence of positing relations without relata.

So suppose we accept that the partial/observables approach is plausibly understood as a form of ontic structural realism: what kind of structure are we dealing with here?  What is the \emph{nature} of the relation between two partial observables, as encoded in a complete observable? I will first look at this question in the context of classical GR, and then consider what changes if we move to a quantum context.

\subsection{Classical GR \label{classical}}

Consider the following example of a classical complete observable, offered by \cite{Rovelli_2002}. Let $\alpha_V$ be the angle that Venus makes with the horizon, and let $\alpha_S$ be the angle that the Sun makes with the horizon. We cannot predict either $\alpha_V$ or $\alpha_S$ individually - Venus makes many different angles with the horizon at different times, so we cannot predict a particular angle without specifying a reference event to establish which time we are interested in, and the same is true for the Sun. But we can predict $\alpha_V$ as a function of $\alpha_S$, i.e. we can write an expression $\alpha_V(\alpha_S) = f(\alpha_S)$ and then we can substitute in some particular value of $\alpha_S$ to find the corresponding value of $\alpha_V$. \cite{Rovelli_2002} tells us that in this scenario, each possible pairing of a value for $\alpha_V$ and the corresponding value for $\alpha_S$ is a complete observable: for example, `(a) complete observable is the value of $\alpha_V$ for $\alpha_S = 5$.' 

Now, one might at first be tempted to understand this partial observable as expressing some kind of spatiotemporal relation between Venus and the Sun. For example, perhaps for each point $s$ on the trajectory of the Sun we are supposed to find the point $v$ on the trajectory of Venus which minimizes the spacetime interval between $s$ and $v$, and then the function $\alpha_S(\alpha_V)$ gives the value of $\alpha_V$  at $v$ when $\alpha_S$ has a certain value at the corresponding point $v$. In that case, the correlations between Venus and the Sun could then be understood as simply a description of regularities occurring in an ensemble of actual events of the form `$\alpha_S = x$ and, at the corresponding closest point, $\alpha_V = y$,' so we could say that we have reduced claims about correlations to claims about an occurrent event or an ensemble of occurrent events. This would suggest that the type of structure relevant here is simply some kind of non-modal, spatiotemporal structure. 

However, it should be noted that the relation between the `correlation' and the actual events is quite different in this case as compared to  pre-relativistic physics. In a pre-relativistic setting, we could  write down the trajectory of the Sun and then separately write down the trajectory of Venus, and then look at these trajectories in order to come up with  a series of coincidence events characterizing how the trajectories happen to be related, and then  finally we could write down a function like  $\alpha_V(\alpha_S)$ encoding the relations we have discovered. Thus the function $\alpha_V(\alpha_S)$ could indeed be thought of as merely a \emph{description} of how the trajectories happen to be related; the correlation supervenes on the underlying events. 

But in the diffeomorphism-invariant case we cannot write down the individual trajectories. We have to start from the function $\alpha_V(\alpha_S)$, and then from that we can derive a set of  coincidence events underwritten by this function. It isn't possible to write down the individual trajectories by themselves, since they would not be diffeomorphism-invariant. So even if the correlation is ultimately interpreted as corresponding to a series of actually-occurring coincidence events, we can't think of the correlation as merely  \emph{describing} those events, since the  coincidence events depend on the function $\alpha_V(\alpha_S)$ and cannot be defined prior to it. Thus looking at the way the prediction is actually arrived at leads to a picture in which the `correlation' expressed by $\alpha_V(\alpha_S)$ is ontologically prior to the actual events.  

Thus although in classical GR the connections between partial observables might superficially look as if they are spatiotemporal, in fact they are modal in character. The complete observables encoding these relations tell us about what is possible  and impossible: one partial observable cannot exist or be observed without the other, because the two are simply aspects of the underlying relation between them. And they ground counterfactuals, such as counterfactuals of the form `If $\alpha_V$ = x then $\alpha_S$ = y,' or perhaps `If you observed the sun from the perspective of Venus when $\alpha_V = x$, then you would see $\alpha_S$ = y.' They do so because the actual events exist only in virtue of the underlying correlation as encoded in the complete observable, so it is not possible that we should have $\alpha_V$ = x without also having $\alpha_S$ = y.  So there are good reasons to think that even in the context of classical GR, complete observables should be understood as instantiating modal connections, and thus the partial/complete observables formalism should be understood as a form of modal structural realism.

\subsection{Quantum Gravity \label{quantum}}

Let us now consider what might change when we move to the context of quantum gravity. \cite{Rovelli_2002} explains that what is physically real in quantum gravity  - the `observables' -   `are relative quantities expressing correlations between dynamical variables.' That is, as in the classical case, observables are defined in terms of  correlations. However, at the current state of development, existing quantum theories of gravity such as string theory and quantum loop gravity don't easily allow us to extract physically meaningful observables, so here I will appeal to  relative states as defined in the quantum reference frame (QRF) formalism as  a first approximation to help us understand what  the observables might look like in a quantum theory of gravity. 

 Various different formalisms have been proposed for the study of quantum reference frames, but here I will focus on the `perspective-neutral' approach \citep{2020acop}, which begins from a global perspective-neutral quantum state $\Psi$ constructed via Dirac quantization, in which we first quantize the theory and then impose constraints such as diffeomorphism-invariance to arrive at the `physical Hilbert space' consisting of all of the physically possible states. We can then choose a subsystem of the universe  $S$ and perform transformations  on the state $\Psi$ to put it in the general form  $| 0 \rangle_S \otimes | \phi  \rangle_R $, where $R$ represents the rest of the universe without $S$. These transformations are defined so as to preserve all physically relevant information,  so this process merely amounts to creating a new representation of $\Psi$. The state $| \phi \rangle$ can thus be understood as encoding the internal view of the system $S$, i.e. we arrive at a claim of the general form `the rest of the universe is in the state $| \phi \rangle$ relative to the system $S$ being in the state $| 0 \rangle$.' 
 
One important special case of the QRF approach arises from choosing as a reference frame a system $S$ which has appropriate formal properties to be regarded as a clock, so we end up with observables of the form `the system is in state $| \psi_i \rangle$ relative to the clock reading $t_i$.' As originally suggested by \cite{PhysRevD.27.2885}, it can be shown that relative to an   `ideal clock,' a quantum system will undergo the standard Schr\"{o}dinger evolution \citep{2020cqtd}, and thus we can recover  what looks like ordinary physics on a fixed spacetime background within   the QRF formalism.   

It should be noted that there are various open questions about the QRF program and the Page-Wootters formalism. In particular, in order to get from `The system is in state $| \psi_i \rangle$ relative to the clock reading $t_i$' to some statement about possible observations, we would need to adopt some solution to the measurement problem, or at least make some interpretational commitments \citep{AdlamPW}. However, whatever view we ultimately take on the nature of the quantum state, we do know that at least in non-relativistic quantum mechanics the quantum state of a system mostly behaves like an intrinsic property of that system: it predicts what we will see when we measure that system, and it evolves autonomously when the system is not interacting with anything. So there is a sense in which the correlations encoded in the state $\Psi$ might be thought of as describing or supervening on a sequence of actually-occurring events:  the system is in state $\psi_1$ relative to the clock reading $t_1$, then the system is in state $\psi_2$ relative to the clock reading $t_2$, and so on.  

But again, as in the classical case, these relational facts can only be obtained by taking the global state $\Psi$ and then deriving from it specific predictions for $| \psi_i \rangle$ for various values of $|t_i \rangle$. So as in the classical case,  looking at the way the prediction is actually derived indicates that the `correlation' encoded in the global state $\Psi$ is ontologically prior to the actual events, rather than vice versa. Thus the correlations are not a mere description, and the individual  quantum states $| \psi_i \rangle$ are not really categorical properties, even though they may sometimes look that way: they are relational states defined by their role in a modal relational structure as encoded by the global state $\Psi$.

Moreover, there is an important difference between the classical and quantum cases which strengthens the modal interpretation of these correlations. In classical GR, we can always work within a Lagrangian picture, and thus to restore determinism in the face of diffeomorphism invariance we can simply assert identities at the level of whole histories - that is, as described in section \ref{partial}, we require that histories related by a diffeomorphism  are physically identical. We can then think of the apparent differences between these histories as arising simply from `our freedom to foliate a manifold in many different ways' which `lets us describe one and the same spacetime in terms of very different sequences of three-geometries' \citep{Wallacenew}. Thus we never have to assert identities between individual \emph{configurations} of events on a spacetime slice, and therefore we can still be realists about the apparent spatial arrangements encoded in these configurations, i.e. we can hold on to the perspective-independent reality of manifest spacetiotemporal structure. Thus in the classical case, one might perhaps argue that although it is true that individual trajectories like those of Venus and the Sun can only be defined in relation to each other, nonetheless the relations between them can  still be thought of as  spatiotemporal in nature, as long as we are careful about the kind of spatiotemporal properties and relations we admit into our description. 

But in the quantum case, the standard approach to quantization of gravity blocks this spatiotemporal interpretation of the correlations. That is, there is a standard procedure known as `canonical quantization' which is commonly used  for quantizing theories subject to a constraint; for example, it was employed successfully by Dirac to construct  quantum electrodynamics \citep{dirac2001lectures}. Since general relativity is subject to the constraint of diffeomorphism invariance, it is natural to try to quantize gravity using canonical quantization. But canonical quantization can only be used within a Hamiltonian formulation of the theory, which means that the diffeomorphism constraint must be applied not to whole histories, but to individual states.   Thus if we apply canonical quantization to a theory of gravity,  in order to restore determinism we must assert identities between \emph{states} rather than whole histories: as \cite{Wallacenew} puts it, `the Dirac quantization algorithm requires the quantum state to be invariant under the action of the local symmetry. This forces the symmetry to be interpreted as a mere redescription of the physics, rather than as a physically meaningful transformation.' This means that if  canonical quantization is indeed the right approach to the quantization of gravity,   the restoration of determinism  in quantum gravity will require a much more radical transformation of the ontology than in the classical case. For asserting identities between \emph{states} related by diffeomorphism transformation would require us to accept that the real content of a quantum state is not in the  spatial arrangement of properties that it appears to encode, since that arrangement will be changed by symmetry transformations, but in its underlying diffeomorphism-invariant correlations. That is, in this context it seems we can no longer hold on to the perspective-independent reality of manifest spacetiotemporal structure - only the correlational structure is physically real.  So in the quantum case, there are even stronger reasons to think we must understand the connections between partial observables as ultimately modal rather than spatiotemporal, since it is the underlying modal connections which survive  under the action of the local symmetry.
 
We can see this feature in action if we consider the kind of relativization appearing within the QRF formalism. As noted earlier, in the classical case, individual statements about complete observables of the form `the value of $\alpha_V$ when $\alpha_S = 5$'  may  superficially look like they are describing some spatiotemporal connection between $\alpha_V$ and $\alpha_S$, e.g. something to do with minimizing  spacetime intervals. But in the quantum case it does not appear that there is even a superficial reading of the connections between relational observables which allows us to think of these connections as spatiotemporal in nature. Given a statement in the QRF formalism  like  `the system is in state $| \psi_i \rangle$ relative to the clock reading $t_i$,' it may be tempting to understand the phrase `relative to' as expressing something like `the clock reads $t_1$ at some time $t'$ and the system is in state $\psi_1$ at the same time $t'$' - but as noted by \cite{AdlamPW} this cannot be what `relative to' means, because in this formalism  time is \emph{defined} by the readings on the clock $t_1$, and there is no time coordinate outside of what is defined by the clock. Indeed, if we take the Page-Wootters formalism seriously, it appears that in the quantum setting   `time' - and hence `spacetime' -  only appears when we adopt a quantum reference frame and describe evolution relative to it, so in a sense spacetime itself emerges only relative to an internal view, and thus it would not make sense to imagine that the connections between partial observables are primarily spatiotemporal.
 
 So in the quantum case it seems even clearer that the partial/complete observables formalism and the QRF formalism should be understood as a form of \emph{modal}   structural realism. The `physically real' content encoded in the  complete observables or the perspective-neutral quantum state is a form of fundamental modal structure, and the features which appear to be intrinsic categorical properties, as encoded in the partial observables or relational quantum states, are dependent on this modal structure. These formalisms can therefore offer useful guidance about how to understand the connection between modal and non-modal features of reality in the context of modal ontic structural realism.

 \section{The Emergence of Non-Modal Features\label{modal}}

Since the partial/complete observables formalism and the QRF formalism arise within the context of GR and quantum gravity, they explicitly reject the idea of using spacetime as a fixed background;  thus they can potentially give us useful ideas about how to account for the connection beteen modal and non-modal features of reality without either appealing to quiddities  or giving spacetime a privileged role. So let us now consider more closely what these formalisms say about the nature of the underlying modal structure and the emergence of non-modal features.

\subsection{Modal Structure \label{ms}}

Objections to modal ontic structural realism often focus on the vagueness of the notion of `modal structure.' The most intuitive way to think about modal structure is to say that it is simply \emph{causal} structure \citep{doi:10.1080/02698590903006917}, but this may not be a generalizable solution, since it has often been argued that causation may not feature in our most fundamental theories of physics (see \cite{10.2307/4543833,Field2003-FIECIA,Rovellioriented,Adlam2023-ADLITC-3}). Yet some authors profess to find it puzzling what `modal structure' could be if it is not causal structure. To respond to these concerns, let us consider the nature of the modal structure appearing in the partial/complete observables formalism.   

First, what kind of modality is involved in the connection between two partial observables? One might be tempted to identify it as a form of physical or nomic necessity, on the grounds that the formalism of partial/complete observables is derived from a specific set of laws of nature, such as the laws of GR. But in fact I think this connection is best understood as a form of  \emph{metaphysical} necessity, because by definition partial observables exist only as aspects of the relations encoded in complete observables, and thus a partial observable is what it is  only in virtue of its relation to the other partial observable with respect to which it is defined. That is, in the classical case described in section \ref{classical}, it is  metaphysically necessary that $\alpha_V$ stands to $\alpha_S$ in the relation encoded by the function $\alpha_V(\alpha_S)$, because this function defines what $\alpha_V$ is; it would not be  a partial  observable at all if it could exist by itself, and it would not be the particular partial observable that it is if it were relativized in some other way. Similarly, in the quantum case described in section \ref{quantum}, it is   metaphysically necessary that the relational quantum states stand in the relations encoded in the perspective-neutral state $\Psi$, because they would not be the specific relational states that they are if they stood in different relations. As I will discuss further in section \ref{lawhood}, this means that in some ways the partial/complete observables formalism is quite closely aligned with the dispositional essentialist view of properties.

Of course, we can always conceive of another possible world in which physics is not diffeomorphism-invariant and hence local functions of coordinates are individually physically real - for example, a world governed only by the laws of Newtonian physics. And in such a world, we could identify a variable which looks superficially similar to $\alpha_V$, which would be a non-relational variable whose existence does not depend on relations to other variables. So it is of course not metaphysically necessary that `the angle that Venus makes with the horizon' must always be relativized to $\alpha_S$ as encoded in the function $\alpha_V(\alpha_S)$.  But it \emph{is} metaphysically necessary that the partial observable $\alpha_V$ which is defined by  the function $\alpha_V(\alpha_S)$ must always be relativized to $\alpha_S$ as encoded in the function $\alpha_V(\alpha_S)$! This example indicates that we must be careful about asserting cross-world identities with respect to partial observables: they  may look like categorical properties in some contexts, but they are not really categorical properties, and thus we cannot straightforwardly identify them with categorical properties in other worlds, nor even with partial observables relativized in different ways. Partial observables are what they are in virtue of their place in some particular relational  structure -  it seems natural that partial observables ought to be defined in structuralist terms, in accordance with  their intrinsically relational nature, and thus the structure which defines them imposes metaphysically necessary relations between them.

Now, at this point one might object that the laws of general relativity are compatible with many different possible worlds, each containing a different assortment of complete observables. So it is both logically and physically possible that different complete observables could have been instantiated in our actual world, and therefore the specific form taken by the  structure encoded in the complete observables in our world is a contingent matter. And surely it cannot be the case that  this structure expresses  relations of metaphysical necessitation if the structure itself is contingent! However, we can reconcile these two perspectives by noting that it is true that the world could have had different relational structure, but in that case the partial observables would have been different; the partial observables that exist in our actual world  could not have existed in a world with different modal structure. That is, it is not metaphysically necessary that these particular partial observables should exist, but given that they exist, it is metaphysically necessary that they stand in the relations that they do.

\subsection{Internal Views} 

The partial/complete observables formalism - and, for similar reasons, the QRF formalism -  appears to instantiate a robust form of modal ontic structural realism in which the underlying modal structure is fundamental and the non-modal features emerge from this underlying holistic structure. Therefore these formalisms offer a useful illustration of what it might look like for non-modal features to emerge from a fundamentally modal reality, so they can help address ongoing concerns about the feasibility of  actual properties and relations emerging from a purely  modal underlying reality.

 In particular, both the total/partial observables formalism and the QRF formalism suggest that  our ordinary perspective on  the world emerges out of a diffeomorphism-invariant theory in virtue of our having an \emph{internal view}  on reality, relative to which partial observables or relative states appear as elements of physical reality. And the methodology of adopting an internal view provides a powerful apparatus for understanding the emergence of  features of our experience which may not appear   in an absolute, observer-independent description of the world.

In  a sense the methodology of adopting an `internal view' is not particularly new. However,  in theories which postulate an absolute spacetime, there is no strong distinction between external and internal views. We can (in principle) describe the universe as a whole from a third-person point of view, and then if we focus on how the world appears to an internal observer, in effect their internal view   corresponds to simply cutting out a small piece of the external description corresponding to the region which is perceptually accessible to the observer, and perhaps putting it in some state of motion. Switching to the internal view does not introduce anything particularly novel to the description. 

But when we move away from an absolute spacetime, things change. For example, if we understand Newtonian spacetime in  relational terms, then the `third-person' description is \emph{purely} relational: it will not contain `positions,' only `relative distances,' and likewise it will not contain `velocities,' only `relative velocities.' Whereas adopting an internal view amounts to choosing a frame of reference relative to which we can define positions and velocities.  So the internal view no longer amounts to simply cutting out a small piece of the third-person description; the internal view \emph{adds} something that was not there in the external view, since it introduces positions and velocities.  In effect, moving from the third-person view to the internal view brings new kinds of  properties into being, since  objects now have positions and velocities and in the formal description enabled by the frame of reference these properties appear to belong to them intrinsically,  whereas in the third-person picture objects only stood in relations of relative distance and relative velocity\footnote{Of course this `bringing into being' is not a real process which happens in time; no one literally moves from the third-person into the first-person perspective. I use this term to emphasize the fact that there is something novel in the internal description which is not present in the external description, which we can illustrate by switching mathematically from a third-person to a first-person description, thus `bringing into being' new features of the mathematical representation which correspond to categorical properties.}. 

Now, in a relational Newtonian spacetime, this behaviour is confined to specifically spatiotemporal properties; other prototypical `categorical' properties like mass, charge and so on are still observer-independent features which look the same in a third-person description and in an internal view. But the partial/complete observables formalism for diffeomorphism-invariant theories seems to implement a similar mechanism for other properties beyond just the spatiotemporal ones. For example, consider a particle whose spin is precessing over time. We would naturally think of `spin' as an intrinsic property of the particle; but in a diffeomorphism-invariant theory, the spin of the particle is only a partial observable, so it is not physically real and does not appear in a third-person description. The third-person description contains  only \emph{complete} observables,  like `the spin of the particle is $s$ relative to the clock reading $t$.' However, when we adopt an appropriate internal view - for example, one associated with the clock - then within the formal description enabled by this choice of reference frame, the spin again appears as an absolute property of an individual particle, rather than a relational property. So moving from the third-person view to the internal view brings new kinds of properties into being: in the internal view objects have categorical properties which appear to belong to them intrinsically, whereas in the third-person view they can only stand in (modal) relations. 

The idea that partial observables represent new kinds of properties which are brought into being by adopting an internal view is made explicit in the  mathematical formalism  used to represent them. At a mathematical level, we begin by defining partial observables on a configuration space, and we then demand that physically real states should obey constraints like diffeomorphism-invariance, thus yielding  an associated reduced phase space on which the complete observables live. This means that the partial observables correspond to extra `unphysical' degrees of freedom \citep{pittphilsci4223}, and those extra degrees of freedom are in effect new properties brought into being by adopting an internal view. For example, in the spin case, we would be able to identify the spin of the particle as a variable living in the configuration space on which partial observables are defined, but there would be no `spin' in the reduced phase space - the property of spin exists only within an internal view. 

We can also see the emergence of categorical properties quite clearly within the QRF formalism. In this picture there do not exist any non-relational intrinsic properties: there is a perspective-neutral global state $\Psi$, but it simply encodes the collection of relational states, so no part of it can be read as playing a straightforward representational role and it does not  encode categorical properties in the usual sense of that term.  However, when we adopt an internal view by taking up  the reference frame associated with a system $S$, we can assign quantum states to other systems, relativized to $S$, and these quantum states behave just like  ordinary physical states, in the sense that within the perspective associated with $S$ they appear intrinsic  to the systems to which they are assigned, and they undergo time evolution relative to $S$ according to the Schr\"{o}dinger equation. Depending on one's approach to the interpretation of quantum mechanics one might understand these quantum states as categorical properties themselves, or as furnishing probability distributions over categorical properties, but either way they clearly encode some information about what we ordinarily think of as categorical properties. Thus by by adopting this internal view we have in effect brought some quantum states  and hence categorical properties into being,   although they exist only within this particular internal perspective. 

So the methodology of adopting an internal view provides a straightforward answer to the question of how  non-modal features can emerge from a fundamentally modal reality. Within these formalisms, from  a third-person point of view there exists only modal structure, but within an internal view on that modal structure there exist (or at least, there appear to exist) various categorical properties and other such non-modal features. The non-modal features are intrinsically tied to the modal structures and thus are ultimately  defined in structuralist terms, but from  an internal perspective they look and act like the ordinary non-modal categorical properties that we are familiar with. The change in perspective associated with going from the third-person description to the internal view explains how it is that something  can ultimately be modal and structural in nature, and yet look and act like a purely categorical intrinsic property under appropriate circumstances.

This approach may also help address some traditional problems for structural realist views. For example, Newman famously objected to Russell's structural realism on the grounds that structural claims fix almost nothing about the unobservable parts of the world \citep{newman1928mr}. That is, if we characterize structure in terms of Ramsey sentences, constructed by dividing the entities posited by the theory into a class of observable entities and a class of unobservable entities  and then quantifying over the unobservable ones, then it can be shown that claims about structure assert nothing about the unobservable world other than the number of entities in it, because as long as we have the right number of entities we can always make them satisfy the given structure by simply choosing extensional definitions for the relations which fit the desired structure. Modern structural realists typically characterize structure in terms of models rather than Ramsey sentences, but one might nonetheless worry that similar objects will apply, since as long as we have enough entities we can define relations to make them fit the relevant model.  It has been  argued \citep{10.1093/bjps/axl020} that adopting specifically \emph{modal} structure addresses this problem, since interpreted relations cannot simply be given extensional definitions to make them fit any set of entities; but \cite{pittphilsci14116} dissents on the grounds that focusing on modality amounts only to extending our commitment to the empirical adequacy of the theory to other possible world or situations, so we  end up collapsing into a form of modal empiricism. 

By contrast, in the approach taken here, our observations are what they are only in virtue of the place they occupy in a complex  global network of modal relations, and thus it is not possible to reduce the modal structure to something like a set of possible empirical observations: we cannot  specify  the observations across possible worlds without also importing the structures which define them. Indeed, it is not even possible to put such a view in the Ramsey sentence form, because we can't separate out the observable parts from the unobservable parts in order to quantify over the unobservable parts, since the observable parts are essentially tied to their place in the structure. So modal structural realism of this kind cannot possibly collapse to modal empiricism, and therefore this kind of approach looks like a promising way to avoid Newman-style objections; of course there may be other related issues in this vicinity, but certainly the Newman objection in its standard form cannot be posed.

\subsection{What are Internal Views Associated With? \label{what}}

One might worry that since the emergence of partial observables appears to depend on the adoption of some internal view,  this  way of understanding the connection between modal and non-modal features introduces an ineliminable subjectivity into science, or gives observers a privileged place in our metaphysics. However, this need not be the case. The fact that we must refer explicitly to a first-person perspective in order to understand the emergence of   features of  the  experience of some observer does not mean that this observer must have a privileged role in our description of reality: it is just that we cannot describe the perspective of the observer without referring to the observer. Van Fraassen makes a similar point in discussing the problem of locating oneself on a map, noting that `Something more than what is contained in the printed map ... is needed. But what is this `more'? Not a mysteriously different sort of fact which cannot be encoded on a map! The scientific story can be complete in the sense of describing all the facts, including that someone does or does not have the `extra' needed for him or her to draw on a particular bit of science. It is just that \emph{describing the having of it} is no substitute for \emph{the having}' \citep{van2008scientific}\index{van Fraassen, Bas}.

 In addition, it should be emphasized that in general, an `internal view' does not have to be associated with a human or any other conscious entity. Physicists working in the QRF formalism sometimes describe the process of shifting into the reference frame of a certain system as `jumping into the perspective of that system,' but this cannot be taken too literally: the formalism simply defines the state of one system relative to another system, and  even if the latter system happens to be an observer, in and of itself the formalism says nothing about what might be \emph{experienced} by that observer. For example, note that you can always choose to define the  velocity of a baseball relative to the location of my physical body, but of course that description alone does not entail that I will see the baseball moving at that velocity - I might have my eyes shut. One might think that the description can be understood in counterfactual terms, describing what I \emph{would} see if I were looking at the baseball,  but this will not work either: for we could similarly describe the velocity of an electron relative to the position of my center of mass, but clearly this representation does not encode any counterfactuals about what I would see if I looked at the electron, since it is too small for me to see with my ordinary perceptual apparatus.  This indicates that partial observables and relative states in the QRF formalism do not in and of themselves have anything to do with conscious experience; they are simply relative descriptions, which may or may not correspond to the experience of a  conscious observer.

  Of course, if  we are  interested in understanding how conscious observers like us come to have experiences \emph{as of} categorical properties and entities, it may  be important at some point to have some story to tell about how a particular internal view can ultimately come to feature in  a conscious experience. And it should be emphasized that this is a significant technical and philosophical problem - in order to tell this story in the context of a theory of quantum gravity  we would first need a complete formulation of the theory, and then we would need to adopt some solution to the measurement problem, and possibly also a solution to the hard problem of consciousness. In particular, if we are using unitary quantum mechanics with no additional structure and no background spacetime, there is a preferred basis problem which seems to result in the appearance of a large number of different consciousnesses inhabiting completely disconnected realities \citep{Stoica_2025}. And 
in the context of the quantum reference frame programme specifically, the preferred basis problem takes the form of the `clock ambiguity,' referring to the fact that different choices of clocks will produce radically different evolution for the rest of the universe \citep{Albrecht_2008}. This is particularly troubling because different choices of clocks will produce the appearance of different evolution laws, so this feature would  seem to have catastrophic epistemic consequences - if we really believed that the world were like this, we would have no reason to believe that the regularities we see correspond to the `right' underlying laws, so the view ends up looking self-undermining - this could be seen as an instance of empirical incoherence, as discussed by  \cite{barrett1999quantum} and \cite{Huggett_2013}\footnote{Thanks to an anonymous reviewer for bringing this up}.

  However, these specific technical issues lie somewhat outside of the remit of the current project. My goal here is not to argue for the correctness and completeness of these specific formalisms, but to see them as works-in-progress which may, though yet unfinished, still yield important conceptual insights into modality and structure in the context of modern spacetime physics. In particular, even if these formalisms ultimately require significant modifications in order to address the preferred basis problem, nonetheless the fact that relational observables are needed even in classical GR provides at least some reason to think some aspects of these approaches are likely to persist in the long run. In any case, as noted  in section \ref{methodology} I think that even if the formalisms themselves are ultimately wrong, nonetheless the idea of categorical properties emerging relative to internal views is a useful conceptual insight which may extend behind these specific formalisms.  And even if we can't yet say exactly which internal views are associated with conscious experience and how they come to be so associated, nonetheless clearly conscious experience \emph{is} in general associated with some kind of internal view, and thus the observation that the adoption of  an internal view brings into being categorical properties and entities yields at least an outline of a story about how observers like us come to experience categorical properties and entities.

\section{Lawhood \label{lawhood}}
 
Let us finish by considering how the view described here fits in the spectrum of philosophical accounts of lawhood. 

Given the centrality of modal structure in this account, it may not be compatible with a Humean approach to lawhood, but it has commonalities with both the dispositional essentialist view and the constraint-based view. First, note that the account of properties suggested by the partial/complete observables formalism, as described in section \ref{ms}, is very much in line with a structuralist view of properties - it suggests that properties  stand in the modal relations that they do as a matter of metaphysical necessity, and thus it shares some DNA with the view that properties have dispositional essences. However, the dispositional essentialist typically conceives of dispositions as causal properties belonging individually to specific concrete entities, which is why there is a need to appeal to spacetime, both to individuate the entities prior to the assignation of the properties, and to provide a temporal direction which can be used to make sense of causation. By contrast, in the partial/complete observables formalism we instead have one big interconnected global structure, which is modal but not strictly speaking causal, with properties appearing from within the point of view of various internal views on that structure. This global way of defining properties makes the essentialist approach more consistent with modern spacetime physics.  

Meanwhile, the holistic, non-causal nature of this modal structure makes it clear that this view also shares some DNA with constraint-based approaches to lawhood, since such approaches conceive of laws as applying all at once to the whole of history in a non-causal, atemporal way. Both the Einstein equations of GR and the Wheeler-DeWitt equation of quantum gravity are most naturally understood as global constraints of this kind. So interestingly, there is a sense in which the essentialist view and the constraint-based view are converging in this picture: it tells us that properties are what they are in virtue of the way they are situated within the modal structure dictated by global constraints. 

Moreover, the approach suggested here may help us resolve some outstanding problems for the constraint-based approach. 
  \cite{adlam2021laws}  defines constraints extensionally as sets of Humean mosaics, but this seems to involve a commitment both to quiddities and to a specially privileged role for spacetime as the background on which the mosaic is defined. But the formalisms discussed here show a different possibility: in these pictures the primary role of the laws of nature is to place constraints on modal structure itself. For example,  in General Relativity, the actual complete observables instantiated in our world must be compatible with the Einstein Equations, and in a putative theory of quantum gravity using Dirac quantization, the actual perspective-neutral global quantum state must be a solution to the Wheeler-DeWitt equation. In these cases, the laws constrain the modal structure without dictating it uniquely - for example, there are many different  solutions to the Einstein equations, and presumably many different solutions to the Wheeler-DeWitt equation. So these laws operate as constraints in a very similar way to the constraints employed  in the formalisms posited by  \cite{adlam2021laws} and \cite{chen2021governing}: the laws define a space of possibilities,  which in this case is a space composed of possible modal structures rather than a space of possible Humean mosaics, and then  one of these possibilities is  selected and actualized as our actual world.    

Now,  in the approaches suggested by \cite{adlam2021laws} and \cite{chen2021governing} the constraints apply directly to the non-modal content, so the selection and actualization of one of the nomic possibilities is the end of the story. But in the case where the constraints apply to modal structure, there is a further step before we arrive at distributions of (apparently) non-modal properties as perceived by realistic observers. And as we have seen in this article, one way to achieve this is to understand the   emergence of the non-modal properties from the modal structure as corresponding to the adoption of an internal view on this modal structure. So in this picture the laws do not apply directly to the non-modal content, but nonetheless the observed distributions of non-modal content are constrained by the laws, because the  possible distributions of non-modal content are circumscribed by the available internal views on the modal structure, which is constrained by the laws. Moreover, since properties associated with partial observables are what they are only in virtue of the way in which they relate to one another within this nomically-determined modal structure, it is metaphysically necessary that they only appear in the distributions permitted by that structure and thus in the distributions permitted by the laws which constrain the possibilities for that structure, so we can understand the action of the constraints with respect to non-modal features as a form of metaphysical necessitation. 
 
Further work is required to understand the consequences of this approach for the philosophy of lawhood more generally, but there are interesting possibilities to use such an approach to help address some longstanding difficulties in this area. Proponents of governance approaches to lawhood have long criticized Humeans on the grounds that the idea that laws merely \emph{describe} distributions of properties or entities seems to undermine the central explanatory significance of laws \citep{Maudlin2007-MAUTMW,doi:10.1093/bjps/axx020}; and  on the other hand, Humeans have often criticized governance approaches on the grounds that  it is hard to understand how laws could possibly \emph{compel} properties or entities to be instantiated in certain ways \citep{Lewis1983-LEWNWF}. So   either on the descriptive view or the governance view, there are open questions about how laws could actually explain anything. 

And in fact, it seems likely that these difficulties are related to the fact that many approaches to lawhood (including both descriptive approaches and  governance approaches), conceive of modal features as standing in contingent relations to the non-modal features.  Whereas if we understand non-modal features in structuralist terms, so we see modal and non-modal features as aspects of one another or we think of non-modal features as arising from modal ones, it is much less mysterious why the non-modal features must be connected in this way, since then the non-modal features must obey certain modal constraints simply in virtue of being what they are. In particular, if we think of categorical properties as simply views on modal structure, then there is no mystery  as to why the properties  must be instantiated in accordance with this structure: the properties are nothing other than `scattered reflections' \citep{Ismael2020-ISMQHN-2} of this underlying structure, so of course they are constrained by that structure. Thus this way of understanding the connection between modal and non-modal features of reality may offer insight into long-standing debates over the explanatory nature of laws. 
 
To conclude, let me note some interesting directions for further investigation. This paper has focused on the emergence of non-modal structure from modal structure in the context of a specific kind of theoretical framework, i.e. a diffeomorphism-invariant spacetime theory. However, it is possible that the `internal view' approach could be expanded beyond this specific theoretical framework\footnote{Thanks to an anonymous reviewer for these suggestions!}. In particular, note that a number of authors \cite{vanzella2024framebundleformulationquantumreference,kabel2024identificationpointlessquantumreference} have recently offered formulations of the QRF approach in terms of fibre bundles - for example \cite{vanzella2024framebundleformulationquantumreference} give a reformulation of the  QRF formalism in which reference frames are smooth local sections of the bundle of (pseudo-)orthonormal frames of the relevant manifold and metric. This allows a geometric treatment of quantum reference frames, which in particular makes it possible to describe them without appealing to coordinate systems. A formalization such as this could potentially be used to clarify and make rigorous the use of `internal views' that I have described in this article. In addition, since the fibre bundle formalism is quite general, the existence of the fibre bundle formulation of QRFs offers hope that the approach taken in this article might also be applicable to other kinds of theoretical frameworks, provided that they can be expressed in a sufficiently similar fibre bundle formulation. 

Alternatively, note that in the context of category theory, the Yoneda lemma implies  that objects may emerge from the `morphisms' (a generalization of relations) \citep{yokura2017remarkyonedaslemma}, which suggests that category theory might also offer a way of generalizing the approach taken here to other theoretical frameworks. The partial/complete observable approach and the QRF formalism as used here are a useful starting point because they come provided with a relatively intuitive account of partial observables as quantities `to which we can associate a measuring procedure leading to a number,' whereas the more abstract approach based on fibre bundles or category theory is less physically transparent, but using the partial/complete observables formalism as an example could perhaps help us see how to make such an approach work more generally.

Second, note that as pointed out by \cite{Schaffer2005-SCHQK-2}, a number of our best philosophical analyses of concepts such as counterfactuals,  conceivability, and propositions and so on use the possible worlds formalism in a way which seems to require that the relationships between modal and non-modal features of reality are in fact contingent. Thus one might argue, as does \cite{Schaffer2005-SCHQK-2}, that these analyses effectively rule out structuralist accounts of the relationships between modal and non-modal features. However, this way of thinking about the relationship between modal and non-modal features of reality assumes that we can settle this question a priori  by appeal to philosophical analyses, without appeal to any facts about the specific nature of the modal structure of our actual world. By contrast, in this article we have encountered reasons to think that the form taken by the relationship   between modal and non-modal features of reality may depend sensitively on that nature of the modal structure. For example, the idea that non-modal features arise from taking an internal view on modal structure is a natural way of accounting for the connection between modal and non-modal features of reality in a world which instantiates something akin to partial and complete observables as they are understood in  GR, but clearly it would not be suitable  for all possible kinds of modal structure - for example, it would not make much sense in the context of a world defined entirely by the laws of Newtonian mechanics. This  suggests that a satisfactory account of the relationship between modal and non-modal features should be informed both by a priori philosophical considerations and also by facts about the nature of the modal structure in our actual world, and thus we should be prepared for the possibility that some philosophical analyses may have to be revised to reflect discoveries in physics giving new insight into the modal structure of our world.   This could be an interesting topic to address in future work. 

\section{Acknowledgements}

Thanks to the participatns of the workshop `Quantum Gravity and the Laws of Nature' (Geneva, December 2024) for very useful comments and discussions. This work was supported by the  John Templeton Foundation Grant ID 63209.

\end{document}